\documentclass[11pt]{article}
\usepackage[a4paper,margin=0.85in]{geometry}
\usepackage[T1]{fontenc}
\usepackage[utf8]{inputenc}
\usepackage{amsmath,amssymb,booktabs,graphicx,siunitx,xcolor,hyperref}
\usepackage[sort&compress,numbers]{natbib}
\usepackage{caption}
\usepackage{subcaption}
\hypersetup{colorlinks=true,linkcolor=black,citecolor=black,urlcolor=black}

\newcommand{\kb}{k_{\rm B}}

\title{\bfseries Classical Reversible Computation by Quantum Coherence}
\author{\begin{tabular}{c}Daniel Loss$^{1,2}$\\ $^1$  Quantum Center and Physics Department\\King Fahd University of Petroleum and Minerals (KFUPM)\\Dhahran, Saudi Arabia
\\ $^2$ Department of Physics, University of Basel, Basel, Switzerland
\end{tabular}}
\date{}

\begin{document}
\maketitle

\begin{abstract}

Rising energy demand from data-center and AI applications~\cite{IEA2025} has renewed interest in reversible computation~\cite{Landauer1961,Bennett1973,FredkinToffoli1982}, where logic need not dissipate heat at every step if information is uncomputed. Implementations have so far been classical: adiabatic CMOS recovers part of the switching energy but still moves thousands of $k_BT$ per logic node at room temperature~\cite{FrankEdwards2025}. Here we propose classical reversible logic implemented 
by coherent  spin dynamics
in a  spin quantum-dot array~\cite{LossDiVincenzo1998}, with inputs and outputs in classical basis states and no algorithmic use of superposition.
The same spin stores, transports, and computes, with unitary rotation replacing irreversible switching. The universal building block is an iToffoli gate driven by DC voltage pulses and exchange between hole spins in Ge/SiGe quantum dots~\cite{Hendrickx2020,Hendrickx2024}. Simulations with realistic model parameters~\cite{WangScience2024,Geyer2024} reproduce the Toffoli truth table and yield a testable error landscape. Because shuttling transports the  bit without measurement~\cite{VanRiggelen2024,DeSmet2025}, logic and data movement remain unitary until readout. Millivolt pulses on femtofarad gates with superconducting lines dissipate only dielectric loss and have no thermodynamic floor. With the loss parameters assumed for a demonstrated device the gate energy is $\sim 10^{2}\,k_BT\ln2$ at 4~K; at a design point within reach of existing devices it can fall below the Landauer scale $k_BT\ln2$, five to seven orders of magnitude less than a CMOS Toffoli. The same semiconductor hardware therefore serves both purposes, supporting quantum algorithms~\cite{Burkard2023} when superposition is used and classical reversible logic otherwise.

\end{abstract}

\section{A quantum-coherent mode of classical reversible computing}

Landauer connected heat generation to information erasure, and Bennett showed that computation can in principle be made reversible by retaining and uncomputing intermediate information \cite{Landauer1961,Bennett1973}. The growing energy demand of computation, including data-center and AI applications \cite{IEA2025}, together with device limits on transistor switching energy \cite{Datta2022}, has renewed interest in these bounds. Reversible logic has long been a foundational idea and an engineering program in adiabatic CMOS~\cite{FrankEdwards2025} and in superconducting adiabatic logic~\cite{Takeuchi2017}. We ask whether reversible \emph{classical} logic can instead be implemented by the coherent quantum dynamics of semiconductor spin qubits, while still using only classical basis states as logical inputs and outputs.

\begin{figure}[t]
\centering
\includegraphics[width=0.86\linewidth]{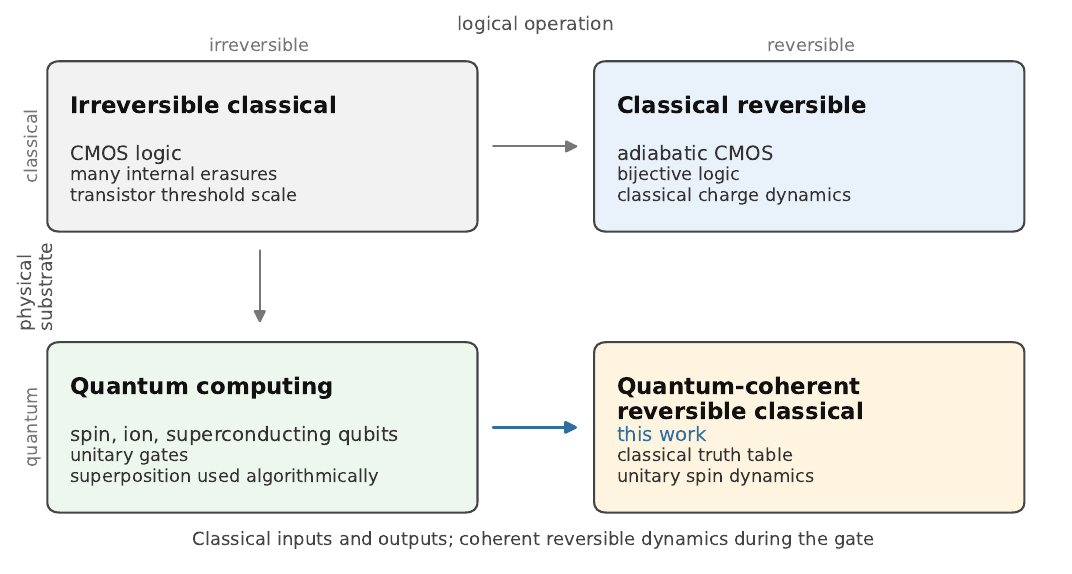}
\caption{\textbf{Device regimes for classical and quantum information processing.} Conventional CMOS (irreversible classical), adiabatic CMOS (classical reversible by quasi-adiabatic charge motion), quantum processors (unitary, using superposition), and the present proposal (classical reversible truth tables by coherent spin dynamics, basis-state inputs and outputs).}
\label{fig:taxonomy}
\end{figure}

This operating mode differs from quantum computing and adiabatic CMOS (Fig.~\ref{fig:taxonomy}). Superposition is not used as a computational resource, so this is not a quantum algorithm. Nor is it classical adiabatic switching: the logic acts on spin rather than charge degrees of freedom. Dissipation arises from applied voltages, decoherence, readout, reset, and information erasure. We refer to this regime as quantum-coherent classical reversible computation.

During storage and local shuttling, the spin bit occupies one of two classical basis states, so the relevant lifetime is the spin-relaxation time $T_1$, not the dephasing time $T_2$. Phase coherence is required only during the gate operation, when the target spin undergoes a controlled unitary rotation. An upper bound on the number of consecutive basis-state operations, set by relaxation alone and independent of gate fidelity, is approximately

\begin{equation}
N_{\rm ops}\sim {T_1\over t_{\rm gate}} .
\label{eq:T1ops}
\end{equation}
Measured hole $T_1$ reaches 32~ms at dilution-fridge temperatures \cite{Lawrie2020,StanoLoss2022}; at 4~K the data are a hole bound of $10~\mu$s \cite{Camenzind2022} and millisecond Si-electron values above 1~K \cite{Yang2020,Petit2018,Huang2024}, and a model calculation for a Ge hole at $0.1$~T out of plane predicts $17$~ms \cite{WangCulcer2021} (SI~\S2). With $t_{\rm gate} = 172$~ns at the Scenario~II operating point (below), the gate supports $6\times10^{2}$--$10^{5}$ operations per $T_1$ at 4~K for an assumed $T_1=0.1$--$17$~ms. Logical depth is also limited by gate and shuttling errors, exchange stability, reset, readout and error correction; the gate-error-limited depth between error-correction cycles, which is much shorter than this relaxation ceiling and is what sets the working circuit depth, is treated in SI~\S\S19, 20 and 27. 

The proposal builds on experimentally established elements in Ge and Si spin-qubit devices, including coherent spin shuttling and baseband spin control \cite{WangScience2024,Hendrickx2024,VanRiggelen2024,DeSmet2025,Unseld2025}. In hopping-based control, a spin is moved between quantum dots with different quantization axes using baseband voltage pulses, avoiding microwave drive \cite{WangScience2024}. Wang \emph{et al.} report single-qubit hopping-gate fidelities near 99.97\% and coherent shuttling fidelities of $\geq 99.992\%$ per hop in a four-dot array \cite{WangScience2024}. Dynamical-decoupling measurements show no measurable spin decay after thousands of shuttles~\cite{WangScience2024}. 
Sweet-spot operation of Ge hole spins further enhances coherence \cite{Hendrickx2024}.
The same spin can therefore compute, carry, and store information (see below and SI~\S\S15, 16): data can be transported by shuttling without readout, so communication remains unitary between projections.

This work introduces the iToffoli gate as a universal building block for quantum-coherent classical reversible computation, driven by all-DC hopping pulses. We analyze its error landscape using measured hopping angles, an exchange coupling within the measured range, and the exchange anisotropy and Zeeman bias as model parameters. We calculate the energy cost. The gate energy is set by dielectric loss and has no thermodynamic floor: it is $\simeq 10^{2}\,k_{\rm B}T\ln 2$ at 4~K with the loss parameters assumed for a demonstrated device and can fall below the Landauer scale $k_{\rm B}T\ln 2$ at a design point within reach of existing devices, five to seven orders of magnitude less than a room-temperature CMOS Toffoli per gate, two to four orders at the wall plug for the device term alone. Details of the Hamiltonian simulations, energy accounting, error-correction estimates and the microscopic Hubbard analysis are given in the Supplementary Information (SI).

\section{Physical model and hopping iToffoli mechanism}

Toffoli, together with NOT and ancilla bits, is universal for reversible Boolean logic \cite{Toffoli1980}. The iToffoli is equivalent to Toffoli for classical inputs because the phase on the activated branch is invisible to $M_z$ (spin-projection) readout. This implementation builds on the framework of gate-defined spin qubits \cite{LossDiVincenzo1998, Burkard2023}. The NOT operation is native in this hardware: with the controls parked (or absent), it is simply a single-spin $\pi$ rotation of the target, demonstrated experimentally~\cite{WangScience2024}. The single-step, resonantly driven (electric-dipole spin resonance, EDSR) iToffoli was proposed~\cite{GullansPetta2019} and demonstrated first on germanium hole spins~\cite{Hendrickx2021}, then on silicon electron spins~\cite{Takeda2022}; both are EDSR-driven. The novelty here lies in the gate mechanism and its energy cost: the same logical gate is driven by DC hopping pulses, not EDSR -- voltage pulses on gate electrodes, as in classical CMOS, with no high-frequency hardware per qubit. EDSR spin flips dissipate substantially more energy than baseband single-spin rotations \cite{WangScience2024}; the quantitative comparison is given below and in the SI.

\begin{figure}[t]
\centering
\begin{subfigure}{0.54\linewidth}
\includegraphics[width=\linewidth]{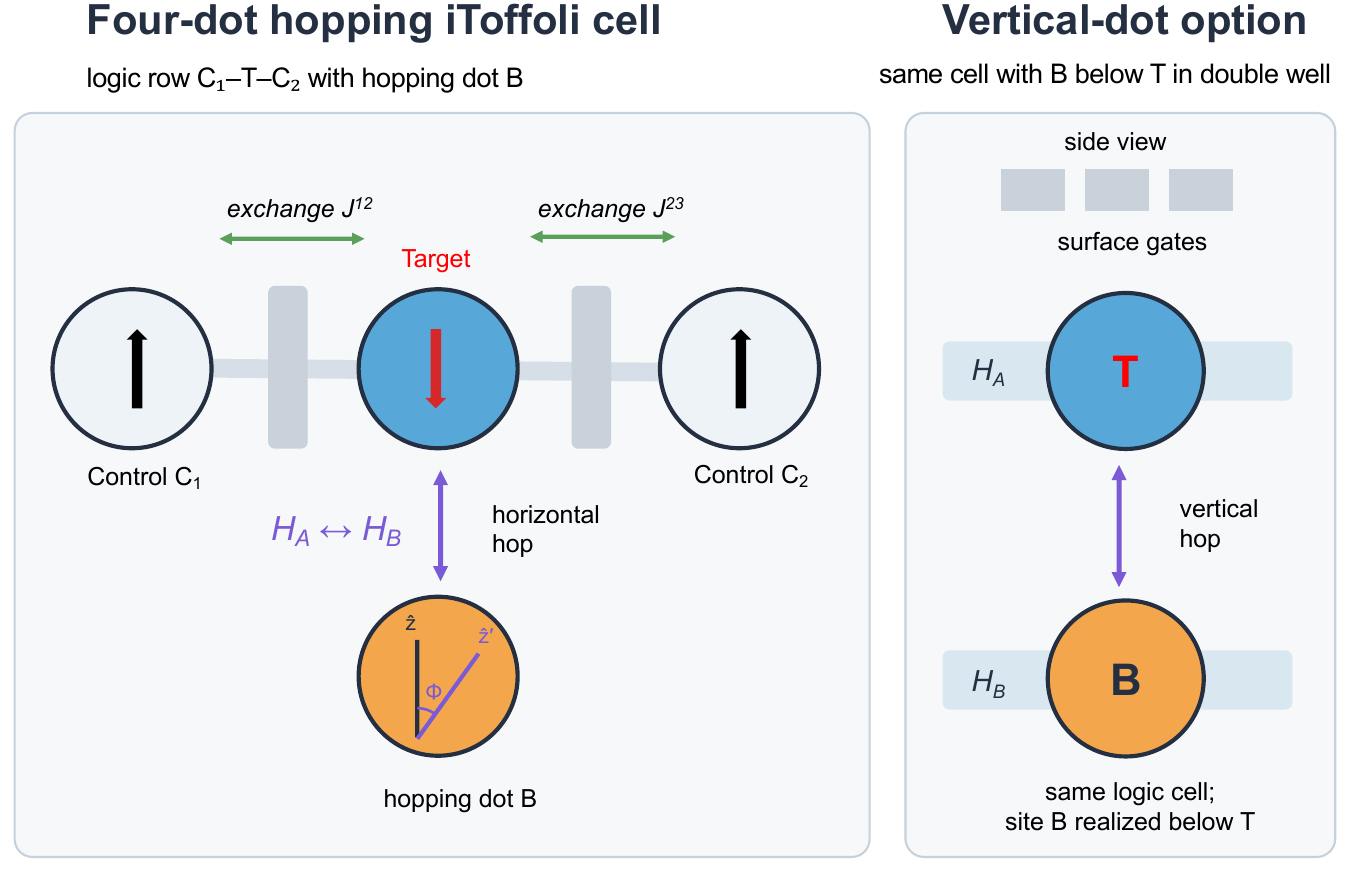}
\caption{}
\end{subfigure}\hfill
\begin{subfigure}{0.46\linewidth}
\includegraphics[width=\linewidth]{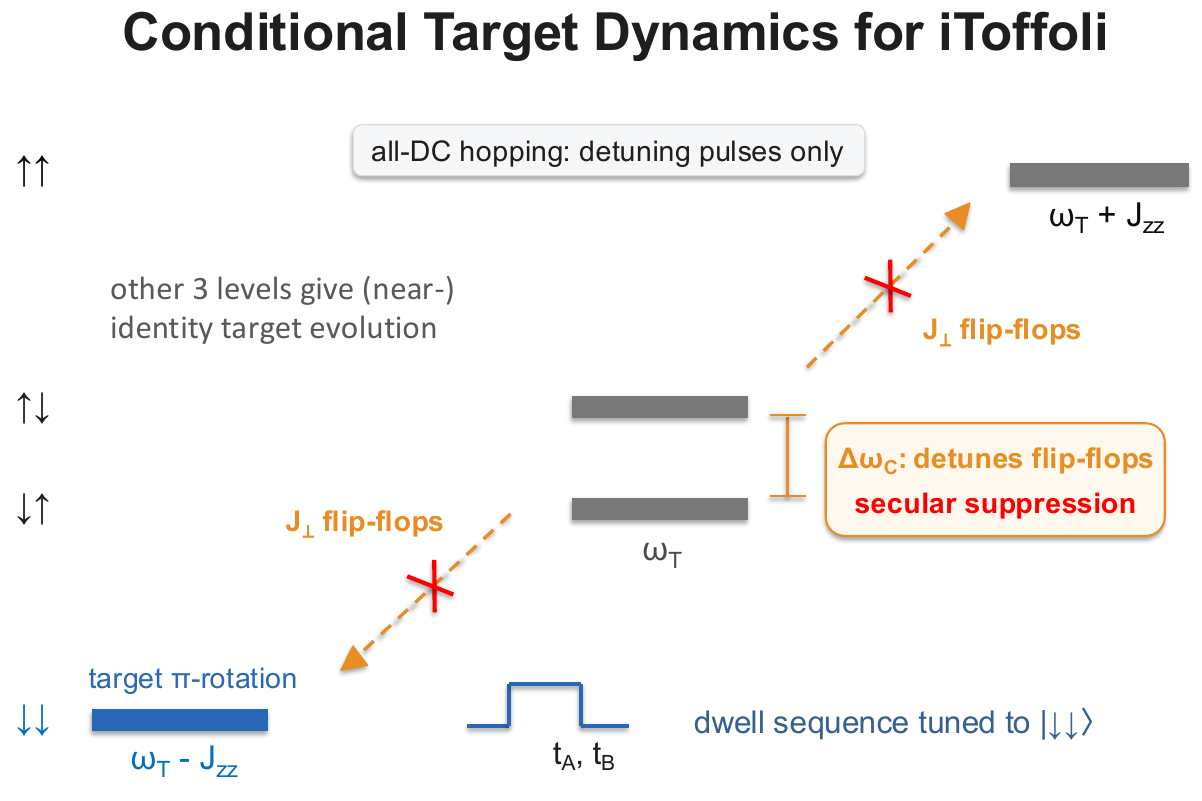}
\caption{}
\end{subfigure}
\caption{\textbf{Hopping iToffoli cell and conditional target dynamics.}
\textbf{a,} Four-dot $C_1$--T--$C_2$ cell: the target spin hops between dots $A$ and $B$ under DC detuning pulses; the two target dots have local quantization axes separated by angle $\Phi$. The auxiliary dot may be lateral or vertical; vertical hole double dots have been demonstrated in strained Ge double wells \cite{Ivlev2024}. Arrows denote charge-adiabatic, spin-diabatic transfers.
\textbf{b,} During dwell $A$ in the Ising limit, the conserved controls define four sectors with target precession frequencies \(\omega_T+J_{zz}\), \(\omega_T\), \(\omega_T\), and \(\omega_T-J_{zz}\). The dwell sequence is chosen so that only the 
\(|\downarrow\downarrow\rangle\)
sector receives an odd-\(\pi\) rotation.  All operations are DC.}
\label{fig:cell}
\end{figure}

The proposed cell is a $C_1$--T--$C_2$ chain, with the target between the two controls (Fig.~\ref{fig:cell}). Placing the target between the two controls keeps every interaction in the gate nearest-neighbor, with no control routed through the target. The target spin hops between two dots, $A$ and $B$, whose spin-quantization axes enclose the angle $\Phi$. We take the axis change experienced by the target spin to be instantaneous (the spin-diabatic limit) and use the calibrated angle $44.7^\circ$ \cite{WangScience2024} as the central operating point; the spin-diabatic approximation and other angle choices are checked in the SI (\S4 and~\S6).

The spin Hamiltonian is written as
\begin{equation}
  {H\over\hbar}=\sum_i \boldsymbol\omega_i\cdot\mathbf s_i+
  \sum_{\langle i,j\rangle}\sum_{\alpha,\beta=x,y,z}
  J_{\alpha\beta}^{ij}s_i^\alpha s_j^\beta,
\label{eq:spin_hamiltonian} %
\end{equation}
where $s_i^\alpha=\sigma_i^\alpha/2$ are dimensionless spin-1/2 operators, and $i,j$ run over the three spins of the cell, i.e. the chain sites $1,2,3$ (left to right), which are the control, target, and control spins $C_1,T,C_2$.
The calculations use the dominant diagonal anisotropy
\begin{equation}
  J_0=\mathrm{diag}(J_\perp,J_\perp,J_{zz}),\qquad r=|J_{zz}|/|J_\perp| ,
\label{eq:Jtensor}
\end{equation}
taken to be the same for both control--target bonds ($J^{12} = J^{23} = J_0$).
The full exchange tensor is decomposed in the SI (\S32); bond-dependent variation is scanned (SI~\S28) and the mismatch between the two dwell configurations is mapped to the gate error (SI~\S33), while Dzyaloshinskii--Moriya and off-diagonal perturbations are listed but not scanned (SI~\S28); the diagonal form used below has the exchange anisotropy $r$ as a model parameter. In hole-spin devices $r$ is set by the spin--orbit interaction, the $g$-tensors and the field direction \cite{Hetenyi2020,BoscoGe2021,Geyer2024,SaezMollejo2025,Massai2026}: it is large in Si fin field-effect transistors \cite{Geyer2024} and close to unity in planar Ge unless tuned \cite{Massai2026} (SI~\S\S1, 9). The gate alternates between two piecewise-constant dwell Hamiltonians $H_A$ and $H_B$, both special cases of Eq.~\eqref{eq:spin_hamiltonian}: the controls have Larmor vectors along $\hat z$ and the exchange on both bonds is the diagonal $J_0$ of Eq.~\eqref{eq:Jtensor}. The only difference is the target Larmor axis, which lies along $\hat z$ during dwell $A$ and tilts to $\hat z\cos\Phi+\hat x\sin\Phi$ during dwell $B$, so the controls stay fixed while the target quantization axis rotates by the hopping angle $\Phi$ on each hop.
Here $\omega_i = 2\pi f_i$ ($i = 1, 2, 3$, with $\omega_2 \equiv \omega_T$) are the Larmor frequencies of the three spins in an external magnetic field of $\sim 25$~mT, with $f_T = 89.57$~MHz the model target value ($g_T=0.256$) at the calibrated operating point \cite{WangScience2024}. The control frequencies $f_1 \simeq 133$~MHz and $f_3 \simeq 46$~MHz are split symmetrically around $f_T$ by the $g$-factor asymmetry, with control Zeeman bias $\Delta\omega_C\equiv\omega_1-\omega_3$ (SI~\S3); they are design values for Scenario~I (the device of Ref.~\cite{WangScience2024} has $42.6$ and $89.5$~MHz at $25$~mT).

In the Ising limit $J_\perp\rightarrow0$, with symmetric bond exchange $J^{12}_{zz}=J^{23}_{zz}\equiv J_{zz}$, the target frequency is shifted by $J_{zz}(s_1+s_3)$ (with $s_i = \pm 1/2$), so only the 
$|\downarrow\downarrow\rangle$
sector matches the chosen dwell sequence. The controls remain in classical eigenstates while the target undergoes the conditional rotation.

\section{Simulation results and microscopic design rules}

The quantum unitary for one hop pair is
\begin{equation}
U_{\rm pair}=\exp(-iH_Bt_B/\hbar)\exp(-iH_At_A/\hbar),\qquad U=U_{\rm pair}^{N} .
\label{eq:gate_unitary} %
\end{equation}

The hop is charge-adiabatic but spin-diabatic: 
the hole follows the lower orbital branch,
whereas the spin does not adiabatically follow the changing local quantization axis.  The dwell
intervals \(t_A,t_B\) are therefore the active precession periods under \(H_A,H_B\).  Spin-adiabatic
shuttling is a different transport mode used when one wants to move a bit without a gate rotation (SI~\S16).

For the angle $\Phi=44.7^\circ$ and $J_{zz}/2\pi=40$ MHz in the Ising limit, numerical optimization of $\varepsilon_{\rm full}$ gives $N=6$, $t_A\simeq17.297$ ns and $t_B\simeq18.897$ ns, so $t_{\rm gate}=217$ ns; the full eight-dimensional simulation reproduces the correct truth table with maximum full-state error $\varepsilon_{\rm full}=0.42\%$. This is an Ising-limit reference point; the realistic operating points used below run faster (Scenario~II, $t_{\rm gate}=172$~ns). Scale-invariance of the Ising protocol under uniform scaling of $\omega_T$ and $J_{zz}$ allows much shorter gate times at higher Larmor frequency (SI~\S11).

For finite anisotropy we use three reference operating points (SI~\S9). The principal point is Scenario~II: Ge/SiGe hole spins with a passive, pre-magnetized Co micromagnet ($r=8$, assumed control Zeeman bias $\Delta\omega_C/2\pi = 960$~MHz, of the size measured with a micromagnet on Si electrons~\cite{Takeda2022}), giving $\varepsilon_{\rm full} = 0.53\%$ at $t_{\rm gate} = 172$~ns, a $\sim 2.1\times$ margin below the $F=81$ majority-vote threshold ($\varepsilon_{\rm full}\le 1.13\%$; $F=3^k$, defined below). This point does not rely on the anisotropy: with isotropic exchange ($r=1$) the same pulse gives $\varepsilon_{\rm full}=0.91\%$. Without a micromagnet (Scenario~I, native Ge/SiGe, $\Delta\omega_C/2\pi = 87.5$~MHz set by the $g$-factor asymmetry of the two controls, a design value tuned onto the resonance) the gate relies on the anisotropy: for $r=8$ it reaches $\varepsilon_{\rm full} = 1.10\%$ by sitting on an Ising-shifted resonance and lies at the threshold edge. A $^{28}$Si variant with micromagnet (Scenario~III, near-isotropic $r\simeq1$) gives $1.09\%$ and is CMOS-compatible, but needs sub-picosecond pulse timing (SI~\S6). The micromagnet is optional only in Scenario~I. The broad margin of Scenario~II comes from $\Delta\omega_C/2\pi\simeq 960$~MHz, which at $25$~mT requires the micromagnet. Without it, same-sign $g$ factors bound the control Zeeman bias by $|\Delta\omega_C|\le 2\omega_T$. Raising $B$ scales $\Delta\omega_C$ and $\omega_T$ together, so with $J_{zz}$ scaled in proportion the gate stays in the threshold-edge Scenario~I window at any field. The micromagnet decouples $\Delta\omega_C$ from $\omega_T$ (SI~\S3). The germanium iToffoli of Ref.~\cite{Hendrickx2021} illustrates this: at $B=1.05$~T the native $g$-spread alone reaches a micromagnet-scale $\Delta\omega_C$ without a micromagnet, yet the gate remains in the Scenario~I regime (SI~\S24). In $^{28}$Si (Scenario~III) the micromagnet is essential: the electron $g$-factor is nearly uniform, so $\Delta\omega_C$ has no intrinsic source, and the near-isotropic exchange ($r\simeq1$) leaves flip-flop suppression entirely to $\Delta\omega_C$. Here, the $F$ are error-correction factors obtained under simple concatenated threefold majority voting (depth $k$, $F=3^k$) (see SI~\S20). They illustrate error-correction overhead, but are not optimized codes. At the Scenario~II operating point the large $\Delta\omega_C$ secularly suppresses flip-flop leakage, and the gate passes $F=81$ across all four reported hopping angles (SI~\S6).

\begin{figure}[t]
\centering
\includegraphics[width=\linewidth]{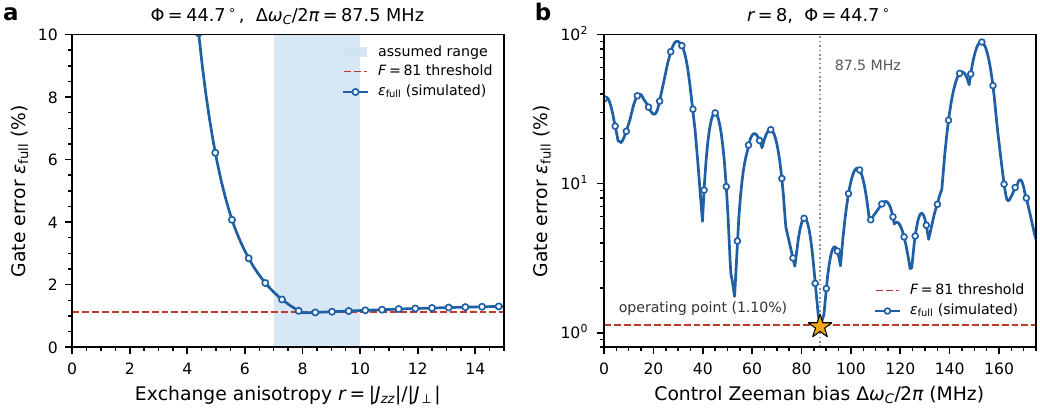}
\vspace{-0.4em}
\caption{\textbf{Error landscape.} \textbf{a,} iToffoli gate error $\varepsilon_{\rm full}$ (maximum over all eight classical inputs) versus exchange anisotropy $r = |J_{zz}|/|J_\perp|$, at fixed control Zeeman bias $\Delta\omega_C/2\pi = 87.5$~MHz and $\Phi = 44.7^\circ$; the gate requires strong anisotropy ($r \gtrsim 8$), and the shaded range $r \simeq 7$--$10$ around $r=8$ brackets the operating point at the $F=81$ threshold ($\varepsilon_{\rm full} \leq 1.13\%$, dashed). \textbf{b,} $\varepsilon_{\rm full}$ versus control Zeeman bias $\Delta\omega_C/2\pi$ at $r=8$; the native operating point ($87.5$~MHz, marked) sits in a sharp local minimum $\varepsilon_{\rm full} = 1.10\%$ on the Ising-shifted resonance. The truth table is correct for all eight classical inputs at the operating point; detailed tables are in the SI (\S\S6, 9) and the exchange-tensor decomposition in \S32.}
\label{fig:main_3_error_landscape}
\end{figure}

Acceptable gate errors are achieved across a finite range of the control Zeeman bias $\Delta\omega_C$ (Fig.~\ref{fig:main_3_error_landscape}), which can in principle be engineered through gate-defined confinement and site-dependent $g$-tensors, measured in hole-spin devices \cite{Hendrickx2024,WangScience2024,Geyer2024,SaezMollejo2025} and attributed in part to strain in the Ge devices. This two-parameter $(r,\Delta\omega_C)$ error landscape is the central prediction.

A separate microscopic concern is exchange stability across the $A\!\leftrightarrow\! B$ hop: $J^A$ and $J^B$, the exchange when the target sits on dot $A$ or $B$, must match to within $\sim 1\%$ for fixed-pulse $F=81$ operation at the Scenario~II operating point. A four-site Hubbard analysis (SI~\S\S29 and 34) shows that this mismatch is set by the bare control--target tunnel amplitudes (device geometry), not by detuning, and that the requirement relaxes to a few percent if the dwell times are recalibrated to the measured exchanges. The vertical auxiliary-dot design changes the target quantization axis while keeping the target laterally aligned with the controls; analogous vertical double-dot geometries with engineered $g$-tensor anisotropy have recently been studied for electron spins in Si/SiGe \cite{SarkarLoss2025}.

The iToffoli is a building block, not a complete processor. Reversible adders and modular arithmetic can be built using standard reversible constructions with Bennett uncomputation \cite{Bennett1973,NielsenChuang2000}, and coherent shuttling forms the local interconnect, as demonstrated in Ge and Si spin-qubit platforms and proposed in recent shuttling architectures \cite{WangScience2024,VanRiggelen2024,DeSmet2025,Kunne2024}. Scaling beyond the iToffoli requires coherent data transport between gates via shuttling, with ancillas that are uncomputed rather than measured. Reversible fanout, ancilla supply, ancilla uncomputation, and CNOT-based routing are analyzed in SI~\S\S16--17.

\section{Energy scale, shuttling, and operating temperature}

The relevant energy for the gate is taken to be the dielectric loss in the gate-electrode capacitance driven by the baseband pulses, as assumed in Ref.~\cite{WangScience2024}, not adiabatic resistive--capacitive (RC) charging.  This holds for superconducting gate lines (SI~\S12). For a single charge transfer (``hop'') with on-device amplitude $\Delta V$, driven capacitance $C_g$, and dielectric loss tangent $\tan\delta$,
\begin{equation}
E_{\rm hop} = C_g\tan\delta\,(\Delta V)^2,
\label{eq:diel_energy}
\end{equation}
where $C_g\tan\delta$ is the loss-weighted capacitance of one driven plunger and each hop is one charge--discharge half-cycle on each of the two plungers of the double dot (SI~\S12). The full per-gate energy is the sum over the $2N=12$ hops of the sequence (SI~\S12), $E_{\rm gate}=12\,E_{\rm hop}$.

With the loss parameters assumed in Ref.~\cite{WangScience2024} ($C_g\tan\delta = 10^{-18}$~F; $\Delta V = 20$~mV, the typical pulse amplitude), $E_{\rm gate}\simeq 4.8\times10^{-21}$~J $\simeq 125\,\kb T\ln2$ at 4~K. This is a reference value, not a floor. Smaller $C_g$, lower $\tan\delta$ and smaller $\Delta V$ are each within reach of existing devices, and together they give $E_{\rm gate}\simeq 0.1$--$0.7\,\kb T\ln2$, the design point of this work (SI~\S12); $0.25$--$1.6\,\kb T\ln2$ if the thermal conditions require the larger detuning, which our estimate favors (SI~\S22). The design-point figures below use the lower band. These figures refer to the twelve-hop, two-parameter pulse used throughout. The same pulse already gives the gate in four hops ($N=2$; $0.52\%$, $57$~ns), reducing the gate energy threefold to $42\,\kb T\ln2$ ($0.03$--$0.2$ at the design point); independently optimized dwell times lower the worst-case error to $0.067\%$ at $108$~ns (SI~\S5).
These are cold-device switching energies for a pulse source that takes the capacitor energy back each cycle (control-electronics budget: SI~\S23); including 4~K refrigeration (coefficient of performance $\mathrm{COP}\sim 3\times10^{-3}$~\cite{Green2019}), the device-level advantage over a $300$~K CMOS Toffoli at the switching floor is $\sim 10^{2}$ at the calibration and $\sim 3\times10^{4}$ at the design point (SI~\S21, \S26); the control electronics add a share per cell that drops below the encoded ($F=81$) device budget for $N_{\parallel}\gtrsim5\times10^{6}$ cells per shared drive source ($1.5\times10^{7}$ against the bare gate; SI~\S23). The refrigeration cost is not fundamental: it scales with the sink temperature, so a colder environment (e.g.\ radiating to the 2.7~K sky in orbit) reduces it.

Local shuttling is one such charge-transfer hop, with the same per-cycle dielectric cost $E_{\rm sh}=E_{\rm hop}\simeq 4.0\times10^{-22}$~J ($10.4\,\kb T\ln2$ at 4 K; SI~\S16).  Present experiments use larger amplitudes optimized for fidelity, but hopping-based control shows a $\sim 20\times$ reduction in dielectric-loss power, estimated from the same loss model, relative to an EDSR benchmark in Si (SI~\S24), with further reduction possible by lowering the pulse amplitudes and increasing the orthogonality of the quantization axes~\cite{WangScience2024}. Data shuttling can therefore remain part of the reversible spin dynamics, without invoking irreversible memory operations.

\begin{figure}[t]
\centering
\includegraphics[width=0.86\linewidth]{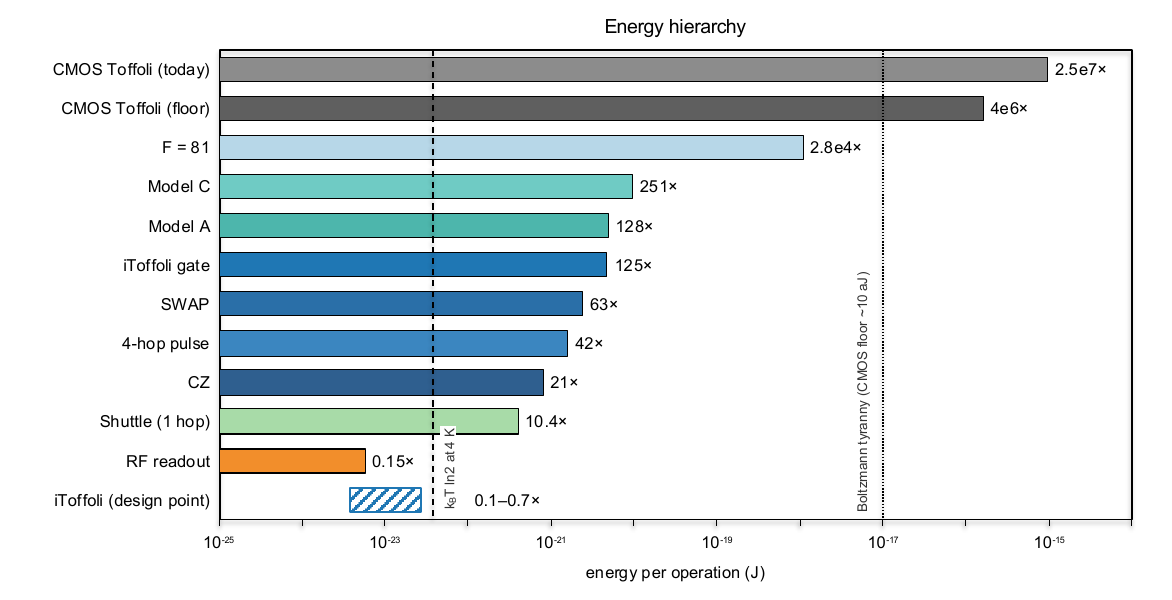}

\caption{\textbf{Energy hierarchy.}
Cold-device energies at 4 K for spin operations. For reference, the dotted vertical line marks the asymptotic $\sim 10$~aJ per-transistor switching floor set by the $V_{\rm dd}\!\geq\!500$~mV `Boltzmann tyranny'~\cite{Datta2022}, and the dark gray bar marks a CMOS Toffoli built from $\sim 16$ such switches ($\sim 1.6\times 10^{-16}$~J in total, SI~\S26); the light gray bar is the same Toffoli at today\textquoteright s circuit-level $\sim 60$~aJ per transistor ($\sim 10^{-15}$~J; our estimate).  Cryo-optimized transistors at $50$~mV are estimated at up to $\sim 30$ times the calibrated gate per Toffoli (SI~\S26).
Spin energies are in units of the 4~K Landauer scale $k_B T \ln 2$; the dielectric-loss iToffoli gate sits at $\simeq 125\,k_B T \ln 2$ for the reference gate stack of Wang \emph{et al.} (one shuttling hop $10.4$, CZ $21$, four-hop pulse $42$, SWAP $63$ for its three CZ pulses) and at $0.1$--$0.7\,k_B T \ln 2$ at the design point (hatched bar; SI~\S12). The readout bar uses Pauli spin blockade for spin-to-charge conversion with dispersive RF detection. The $F=81$ bar folds in both the redundant reversible compute and the irreversible majority-vote (Landauer) cost, the latter charged at $300$~K (SI~\S20). Models~A and~C are end-to-end energy budgets (SI~\S19): Model~A includes readout and reset every cycle ($\sim 128\,k_B T \ln 2$ per gate); Model~C concentrates irreversibility at the output with intermediate uncomputation ($\sim 251\,k_B T \ln 2$ per compute--uncompute pair). Refrigeration overheads (\S21), parameter dependence (\S12), and detector parameters at $\tau_{\rm int}=1\,\mu$s (\S14) are in the SI.}
\label{fig:energy}
\end{figure}

The same exchange that drives the iToffoli also provides
a nearest-neighbor CNOT gate. Two-qubit gates have been demonstrated in these platforms with resonant drive
\cite{Hendrickx2020,Geyer2024,Burkard2023} and, in planar Ge, with baseband exchange pulses~\cite{Hendrickx2021,WangScience2024}; at the  working
point $J_{zz}/2\pi = 40$~MHz, the baseband conditional-phase pulse used here takes
$\pi/J_{zz} = 12.5$~ns and costs $\sim 21\,\kb T\ln 2$ per operation (two hops, one out-and-back charge displacement, for switching the exchange on and off; Fig.~\ref{fig:energy}, SI~\S17); SWAP
follows from three CNOTs by logical decomposition. The same SWAP reorders spins along the line (e.g.\ $C$--$T$--$C \to C$--$C$--$T$), so reconfiguration and routing stay within a single 1D chain rather than requiring a two-dimensional network. The same primitives can compose the Toffoli itself (three controlled-phase plus four single-qubit hopping gates, the relative-phase Margolus construction), but the native single-step iToffoli realizes it as one calibrated waveform rather than seven independently calibrated operations, at comparable energy ($12$ versus $14$ hops) and up to $\sim 4\times$ lower error (SI~\S18).

The energy hierarchy of Fig.~\ref{fig:energy} includes a readout bar for dispersive RF reflectometry~\cite{Vigneau2023,Crippa2019,West2019,Pakkiam2018,Hogg2023}, giving a cold-device dissipation of $\sim 0.15\,\kb T\ln 2$ in the gate line for one $1\,\mu$s probe pulse (the probe delivers $83\,\kb T\ln2$; SI~\S14), three orders of magnitude below the gate at the calibration and of the order of the gate at the design point.  Readout uses Pauli spin blockade for the spin-to-charge conversion~\cite{Ono2002,Petta2005}, and the same detector performs initialization, so the boundary cost is paid twice per logical bit independently of circuit depth.  The Model~A bar in Fig.~\ref{fig:energy} counts readout and reset at the Landauer limit ($3\,\kb T\ln 2$ per cycle); the Model~C bar has no readout term (SI~\S19).

The first experiment need not operate at 4 K. The truth table and error
rates can be tested at sub-Kelvin temperatures, where Ge/SiGe
hole-spin coherence is already established. Hole spin-phonon relaxation falls steeply with decreasing field~\cite{WangCulcer2021}, and an optimal field geometry, predicted for Ge/Si nanowires, suppresses the phonon matrix element further~\cite{MaierKloeffelLoss2013,ScappucciRoute2021}; high-fidelity Ge-hole control above 1~K \cite{Hendrickx2024} and coherent single-hole-spin qubit operation \emph{above} 4~K in a silicon FinFET \cite{Camenzind2022,MaurandJehl2022}, together with electron-spin operation above 1 K \cite{Petit2020,Yang2020}, have already been demonstrated, and hole coherence is expected to improve further with material quality and isotopic purification~\cite{ScappucciRoute2021,FangCulcer2023}. The platform is also scaling rapidly: germanium hole spins have recently been operated as an 18-qubit modular array with simultaneous control and mean (median) single-qubit gate fidelity $99.8\%$ ($99.9\%$)~\cite{Dijkema2026}.

Operation at 4~K is the goal: 4~K stages provide milliwatt-to-watt cooling power, three to four orders of magnitude more than a dilution refrigerator, which allows control electronics next to the array \cite{MaurandJehl2022} (control budget: SI~\S23). The Zeeman field is modest ($\sim 25$~mT) and supplied by a superconducting magnet with no static dissipation; on-chip nanomagnet patterns offer a scalable alternative \cite{Unseld2025}.

For the $172$~ns gate, the dominant charge noise is quasi-static $1/f$ noise, for which the dephasing error scales quadratically as $\varepsilon^* \approx (t_{\rm gate}/T_2^*)^2/3 \simeq 5\times10^{-4}$ at the measured free-induction time $T_2^* \simeq 4.5\,\mu$s in Ge hole-spin qubits~\cite{WangScience2024} (measured below $140$~mK and without a micromagnet), sub-leading to the intrinsic gate error $\varepsilon_{\rm full}$ (SI~\S2).

Compared with CMOS, three differences emerge. The voltage scale: CMOS relies on transistors constrained by the subthreshold slope (the `Boltzmann tyranny') and the threshold voltage \cite{Datta2022,FrankEdwards2025}, placing the practical CMOS gate voltage near $\sim 0.5$~V against the $\sim 20$~mV on-device hopping pulse, a factor $\sim 6\times10^{2}$ in $V^{2}$. The small loss-weighted capacitance and the per-cycle dielectric loss account for the rest: the calibrated gate against a CMOS Toffoli at today\textquoteright s circuit-level $\sim 60$~aJ per switch ($\simeq 10^{-15}$~J; our estimate, SI~\S26) gives a factor $\sim 2\times10^{5}$, and the design point against the asymptotic $16\times10$~aJ floor~\cite{Datta2022} a factor $\sim 10^{7}$ (SI~\S26). Encoding, readout, refrigeration, fanout, routing and memory traffic can reduce this advantage at the system level. The memory architecture: a bit is stored in the same gate-defined quantum dot that performs the logic, with storage time set by spin relaxation $T_1$, so the memory wall \cite{Datta2022}, the off-chip traffic that dominates CMOS energy budgets, is avoided for data resident on the array (SI~\S15). The quantum dot footprint is set by lithography and therefore scalable with the fabrication node~\cite{Zwerver2022,Maurand2016}; quantum-dot gate layers are already patterned by extreme-ultraviolet lithography on 300-mm wafers~\cite{Neyens2024,ImecDiraqEUV2026}, and since $E_{\rm gate}\propto C_g$ the device loss falls with the node. The energy--delay product: with the gate energy fixed by the hop count and independent of speed (SI~\S12), faster operation lowers the energy--delay product rather than trading energy for speed.

The size of this advantage depends on the required reliability. AI inference, part of a rapidly growing data-center workload~\cite{IEA2025}, is error-resilient: such applications accept losses of output quality of up to $\sim 10\%$~\cite{Mittal2016} (SI~\S26). The ratios above hold at the raw gate error ($F=1$). Higher-reliability tasks add majority-vote redundancy, which both technologies pay, leaving, for the calibrated gate against the floor ($3\times10^{4}$ at $F=1$), $\sim 10^{4}$ at matched $F=81$ and $\sim 10^{3}$ against CMOS at its own redundancy, assumed to be $F=3$--$9$ (SI~\S26).

Compared with the exchange-only (EO) spin-qubit encoding~\cite{HRLqpu2026,Madzik2025}, the single-spin iToffoli reaches comparable gate error at about $20\times$ lower per-operation energy on the same loss model, with data movement by single-spin shuttling rather than encoded-SWAP chains (SI~\S25).

\section{Experimental tests and outlook}

The program has six experimental tests: (i) the three-spin $C_1$--T--$C_2$ truth table; (ii) the $\varepsilon_{\rm full}(\Delta\omega_C, r)$ error landscape, especially the flip-flop band at small $\Delta\omega_C$ and recovery in the secular-suppression regime; (iii) exchange stability during hopping, quantified by $J^A\simeq J^B$; (iv) direct measurement of the driven capacitance, of the loss tangent of the gate stack at 4~K and at the pulse frequencies of the gate, and of the shuttling energy; (v) the charge and orbital relaxation times of the target at the dwell detuning at 1--4~K (SI~\S22); and (vi) the hole spin $T_1$ at 4~K. The error landscape and the per-gate energy follow directly from measured pulse amplitudes, capacitances and dwell times. All six can be tested in present Ge/SiGe platforms; (v) fixes the orbital splitting and detuning the 4~K operating point needs, a splitting planar Ge dots already reach~\cite{Sommer2026}. Beyond these device-level tests, SI~\S27 gives a circuit-level blueprint: a one-bit reversible full adder on five quantum dots, with a simulation-verified schedule and an energy, time, and error budget (a hop-count lower bound, $543\,\kb T\ln2$ per pass), sized to fit demonstrated Ge/SiGe arrays.

Confirming these tests would establish a reversible building block whose energy cost has no thermodynamic floor and, at the design point, can fall below the 4~K Landauer scale. The industrial roadmap for semiconductor spin-qubit quantum hardware and the post-CMOS roadmap for reversible classical logic would then be two operating modes of one platform; progress for either would advance both.

\section*{Acknowledgments}
The author thanks Jelena Klinovaja and Rayyan Raza for discussions and acknowledges the Deanship of Research and the Quantum Center at KFUPM for the support received under Grant
no. CUP25102 and no. INQC2600, respectively. 
During the preparation of this work, the author used generative artificial intelligence tools, including Claude and ChatGPT, to assist with mathematical exploration, code generation, numerical analysis, figure preparation, and manuscript development. All AI-generated outputs were independently verified and reviewed by the author. The author retained full responsibility for the scientific content, interpretations, and conclusions of this work.

\section*{Data and code availability}

This is a theoretical study; no experimental datasets were generated or analyzed. The simulation code and the data files reproducing the numerical figures and tables of the main text and Supplementary Information are available on reasonable request and will be deposited at Zenodo \cite{loss_2026_zenodo} under the MIT license upon publication.

\end{document}